\documentclass{aa}
\usepackage{epsfig}
\usepackage{rotating}
\usepackage{natbib}
\usepackage{graphicx}

\newcommand{\chan}{{\sl Chandra}}
\newcommand{\rosat}{{\sl ROSAT}}
\newcommand{\asca}{{\sl ASCA}}
\newcommand{\sax}{{\sl BeppoSax}}

\newcommand{\spitzer}{{\sl Spitzer}}

\newcommand{\nh}{N_{\rm H}}

\newcommand{\hst}{{\sl HST}}
\newcommand{\vlt}{{\sl VLT}}

\newcommand{\ctio}{{\sl CTIO}}
\newcommand{\aao}{{\sl AAO}}

\newcommand{\fors}{{\sl FORS1}}
\newcommand{\naco}{{\sl NACO}}
\newcommand{\nacon}{{\sl NAos COnica}}

\newcommand{\tmass}{{\sl 2MASS}}
\newcommand{\gsc}{{\sl GSC-2}}

\newcommand{\dao}{{\em Daophot}}
\newcommand{\all}{{\em Allframe}}

\begin{document}

 \title{VLT observations of the Central Compact Object in the Vela Jr. supernova remnant\thanks{Based on observations collected at the European Southern Observatory, Paranal, Chile under  programme ID 074.D-0729(A),077.D-0764(A) } }

 \author{R. P. Mignani\inst{1}
 \and 	
 A. De Luca\inst{2}
 \and
 S. Zaggia\inst{3}
 \and
D. Sester\inst{3}
\and
 A. Pellizzoni\inst{2}
 \and
 S. Mereghetti\inst{2}
  \and
 P. A. Caraveo\inst{2}
}

   \offprints{R. P. Mignani}

   \institute{Mullard Space Science Laboratory, University College London, Holmbury St. Mary, Dorking, Surrey, RH5 6NT, UK\\
              \email{rm2@mssl.ucl.ac.uk}
\and
INAF, Istituto di Astrofisica Spaziale, Via Bassini 15, Milan, 20133, Italy \\
 \email{[deluca; sester; alberto; sandro; pat]@iasf-milano.inaf.it}
\and
INAF, Osservatorio Astronomico di Padova, Vicolo dell'Osservatorio 5, Padua, 35122, Italy \\
 \email{simone.zaggia@oapd.inaf.it}
}

     \date{Received ...; accepted ...}

   \abstract  {X-ray  observations  have  unveiled  the  existence  of
enigmatic  point-like sources  at the  center  of young  (a few  kyrs)
supernova remnants.   These sources, known as  Central Compact Objects
(CCOs),  are thought  to be  neutron stars  produced by  the supernova
explosion,  although  their X-ray  phenomenology  makes them  markedly
different  from  all  the  other  young neutron  stars  discovered  so
far.}{The aim of this work is to search for the optical/IR  counterpart of 
the Vela Junior CCO
and  to understand the  nature of  the associated  $H_{\alpha}$ nebula
discovered  by Pellizzoni et  al. (2002).}{We  have used  deep optical
($R$ band) and IR ($J,H,K_s$ bands) observations recently performed by
our group with the ESO \vlt\ to obtain the first deep, high resolution
images of  the field with the  goal of resolving  the nebula structure
and  pinpointing  a point-like  source  possibly  associated with  the
neutron star.}{Our  $R$-band image shows  that both the  nebula's flux
and  its structure  are very  similar to  the $H_{\alpha}$  ones, 
suggesting that the nebula  spectrum is dominated by pure $H_{\alpha}$
line  emission.   However, the  nebula  is  not  detected in  our  IR
observations, whick makes it  impossible to to constrain its spectrum.
A faint  point-like object ($J  \ge 22.6$, $H  \sim 21.6 $,  $K_s \sim
21.4 $)  compatible with the  neutron star's \chan\ X-ray  position is
detected in our  IR images ($H$ and $K_s$) but not  in the optical one
($R \ga  25.6$), where it is  buried by the  nebula background.  }{The
nebula is most likely a  bow-shock produced by the neutron star motion
through the  ISM or, alternatively, a  photo-ionization nebula powered
by  UV radiation  from  a  hot neutron  star.   A synchrotron  nebula,
powered by  the relativistic particle  wind from the neutron  star, is
the  less likely interpretation  because of  its non-detection  in the
X-rays and of the apparent  lack of continuum emission.  The candidate
CCO counterpart could be the neutron star itself, a fallback disk left
over from the supernova explosion, or  a mid M-type star, or later, at
a distance of 2 kpc.  }

             \keywords{Stars: neutron, Stars: individual: CXO J085201.4-461753}

\titlerunning{\vlt\ observations of the Vela Junior CCO}

   \maketitle

\section{Introduction} 

Ground  breaking observations  with  the current  generation of  X-ray
observatories have  recently unveiled unexpected  and puzzling aspects
of the phenomenology of Isolated  Neutron Stars (INSs).  It has become
clear  that not  all  INSs are  born  as fast  spinning, active  radio
pulsars. A  handful of isolated  compact sources, most  probably young
INSs, do not show any ``standard'' pulsar activity.  They are known as
``Central Compact  Objects'' (CCOs)  in supernova remnants  (Pavlov et
al. 2004), enigmatic compact sources lying very close to the center of
young  (0.3-20 kyrs)  supernova remnants,  showing  only thermal-like,
often unpulsed  X-ray emission. They  are supposed to be  the youngest
members of the radio-quiet NS family which also includes the Anomalous
X-ray Pulsars  (AXPs) and the  Soft Gamma-ray Repeaters  (SGRs), which
make up the class of the {\em magnetars} (Woods \& Thompson 2006), the
INSs with X-ray dim thermal emission (XDINSs, Haberl 2005), as well as
high-energy pulsars  like Geminga (Caraveo et al.  1996).  At present,
we cannot even exclude that  such sources represent ``the rule'' among
INSs, radio  pulsars being  ``the exception'', but  much easier  to be
observed.   Their overall  properties are  largely unknown.   First of
all, the lack of observable  radio emission has still to be explained.
It could be possibly due  to an unfavourable beaming, to inhibition by
an ultra-high magnetic  field, as invoked for the  {\em magnetars}, or
to quenching by accretion from a residual disk.  The nature of the CCO
thermal X-ray emission  is also unclear.  It could  be produced by the
cooling NS surface, by hot polar caps, by the NS magnetic field decay,
or by  accretion from  a residual disk.   Their relationship  (if any)
with other classes of radio-quiet INSs is also debated.

As  in the  case of  other radio-quiet  INSs,  optical/IR observations
could provide a major step  forward in the understanding of the nature
of the CCOs.  Sanwal et al. (2002) identified a field M dwarf detected
by \hst\ as a likely IR counterpart to the RCW103 CCO, possibly making
it the  first X-ray Binary  associated to a SNR.   The identification,
however, is still unconfirmed. IR observations of other CCOs have been
recently carried  out (Fesen et al.   2006; Wang et al.  2007), but no
viable candidate  counterpart has been identified.   An IR counterpart
to the  PKS 1209-52 CCO was proposed  by Pavlov et al.  (2004) but the
identification  has been  recently discarded  both by  Mignani  et al.
(2007a) and Wang et al.  (2007) on the basis of its significant offset
with respect to the  \chan\ position.  The unsuccessful identification
scoreis explained  by the fact that  most CCOs are  rather distant and
heavily  absorbed, and therefore  they are  very difficult  targets at
optical/IR wavelengths.

The only possible  exception is the X-ray source at  the center of the
RXJ  0852.0-4622 supernova remnant,  indeed one  of the  least studied
CCOs.  RXJ 0852.0-4622  (G266.1-1.2) is a very young  (a few thousands
years) shell-like supernova remnant  discovered in the \rosat\ All Sky
Survey (Aschenbach 1998), whose  projected position is coincident with
the southeast edge of the  more extended Vela supernova remnant, hence
dubbed ``Vela  Junior''.  However,  its estimated distance  of $\sim$1
kpc (Slane et al.  2001) puts it beyond the Vela supernova remnant and
thus it rules out any intriguing association between the two.  The CCO
in Vela  Jr. (AX  J0851.9-4617.4) was discovered  by \asca\  (Slane et
al. 2001)  and  studied with \sax\  (Mereghetti 2001)  and later
with \chan, which also provided its sub-arcsec position (Pavlov et al.
2001).   The  X-ray emission  of  the  Vela  Jr.  CCO,  hereafter  CXO
J085201.4-461753, is characterized by a thermal-like spectrum with $kT
\approx  404$ eV  ($\nh \approx  3.5 \:  10^{21}$ cm$^{-2}$  )  and an
emitting radius of  $\approx 0.28$ km (at 1 kpc),  with no evidence of
pulsations (Kargaltsev et al.  2002).  First optical investigations of
the  CXO J085201.4-461753  field were  carried  out by  Pavlov et  al.
(2001) with  the  \ctio\ 0.9m  telescope,  and  by  Mereghetti et  al.
(2002) using  archival  observations obtained  with  the ESO/MPG  2.2m
telescope.  The derived  optical flux upper limits of  $B \sim 23$ and
$R \sim  22.5$ (Mereghetti et  al.  2002) imply  a very high  X-ray to
optical flux  ratio which virtually certified  CXO J085201.4-461753 as
the compact remnant of the  Vela Jr. supernova explosion.  Although no
optical  candidate  counterpart  was  detected by  Mereghetti  et  al.
(2002),  a short  $H_{\alpha}$ exposure  unveiled the  presence  of an
extended   emission  feature   ($\sim$   6''  diameter)   positionally
compatible with the CXO J085201.4-461753 error circle.  The reality of
this feature  was confirmed  soon after by  Pellizzoni et  al.  (2002)
using a deeper $H_{\alpha}$ exposure  ($\sim$ 3 hours) taken as a part
of the  Southern Galactic Plane and Magellanic  Clouds digitized plate
survey (Parker \& Phillips 1998) carried out with the \aao\ UK Schmidt
Telescope.    Unfortunately,  the  low   spatial  resolution   of  the
$H_{\alpha}$ image  (1''/pixel) did not  allow to resolve  the feature
structure  and   to  favour  one  of  the   two  alternative  proposed
interpretations i.e., either a  velocity-driven bow-shock nebula, or a
photo-ionization nebula, powered  by UV radiation from a  hot INS, and
therefore to better  constrain the properties of the  CCO.  \\ Here we
present the results  of the first deep optical  and IR observations of
the Vela Jr. CCO recently performed with the ESO \vlt.  Optical and IR
observations  are presented  in \S2  and \S3,  respectively  while the
results are discussed in \S4.

\section{Optical Observations}

\subsection{Observations Description}

We have performed  deep optical observations of the  Vela Jr. CCO with
\fors\  (FOcal  Reducer  Spectrograph),  a multi-mode  instrument  for
imaging and long-slit/multi-object  spectroscopy mounted at the second
Unit  Telescope  (UT2)  of   the  \vlt\  (Paranal  Observatory).   The
observations were carried out in Service Mode on January 15th and 17th
2005  through the  $R$ Bessel  filter ($\lambda=6570  \: \AA  ; \Delta
\lambda= 1500 \:\AA$).  The four port read out mode and high gain were
choosen  as the  default  instrument configuration.   To minimize  the
light  pollution from  ``star Z''  of Pavlov  et al.   (2001), located
$\sim 1\farcs5$ away from our target, we split the integration time in
20 exposures  of 260 s  each.  In order  to achieve the  best possible
spatial   resolution,   useful  to   disantangle   any  point   source
contribution from the PSF wings of star Z, \fors\ was used in its High
Resolution  (HR) mode  ($1 \times  1$ binning),  with  a corresponding
pixel size of 0\farcs1.  The very bright stars HD 76060 ($V=7.88$) and
Wray 16-30 ($V=13.8$), respectively  $\approx$ 35'' and $\approx 25''$
away from our  target (Mereghetti 2001), were masked  using the \fors\
occulting  bars.   The observations  were  collected  with an  average
seeing of  $\sim 0\farcs8 - 0\farcs9$  and an airmass  $\sim 1.3$ (see
Table  \ref{forsdatasummary}) on  both  nights and  with a  fractional
lunar illumination of  57\% and 47 \% for the  first and second night,
respectively.  

\subsection{Data Reduction and Analysis}

Usual reduction  steps to remove instrumental  signatures were applied
through        the       ESO        \fors\        data       reduction
pipeline\footnote{www.eso.org/observing/dfo/quality/FORS1/pipeline}
using  bias frames  and twilight  flat fields.   Then,  single reduced
images  were combined to  filter out  cosmic rays.
Since no  dither was applied  between single  exposures to
preserve  the masking  of  the bright  stars  in the  field, no  image
registration was required before  the frame stacking.  The photometric
calibration was  performed through  the observation of  standard stars
from  the field  PG1323-86  (Landolt 1992),  yielding $R$-band  zero points  of
$26.86 \pm  0.05$ and $27.48  \pm 0.03$
for  the first and  the second
night,  respectively.    We  note  that the  first  night was  not
photometric, with the computed zeropoint deviating by $\sim$
0.6                 magnitudes                 from                the
trend\footnote{www.eso.org/observing/dfo/quality/FORS1/qc/zeropoints}.
Thus,  we  have  re-normalized  our photometry  using  as  a
reference the  zeropoint of the  second night which, instead,  is very
well consistent with the zeropoint trend. 

\begin{table}
\begin{center}
  \caption{Summary of  the \vlt\/\fors\ $R$-band observations of  the Vela Jr.
field with the number of exposures per night, the single exposure time, the average seeing and airmass  values. }
\begin{tabular}{lllll} \hline
yyyy.mm.dd     & N & Exp. (s) & Seeing  & Airm.	\\ \hline
2005.01.15     & 10 & 259 & 0\farcs83 & 1.3  \\
2005.01.17     & 10 & 259 & 0\farcs93 & 1.3  \\  \hline
\end{tabular}
\label{forsdatasummary}
\end{center}
\end{table}

\begin{figure*}
\centering 
\includegraphics[bb=36 150 576 690,width=8.0cm,clip]{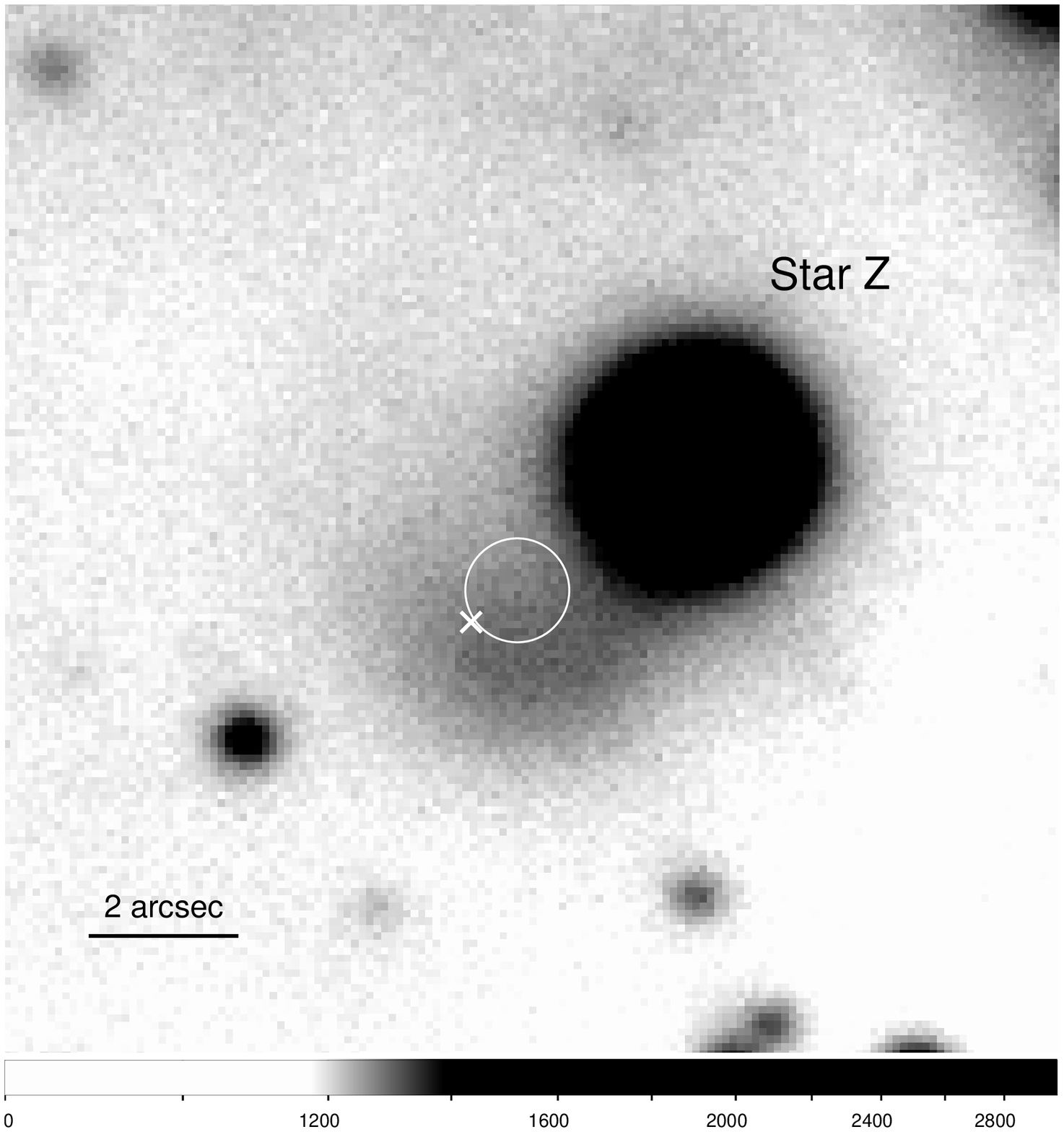}
\includegraphics[bb=36 150 576 690,width=8.0cm,clip]{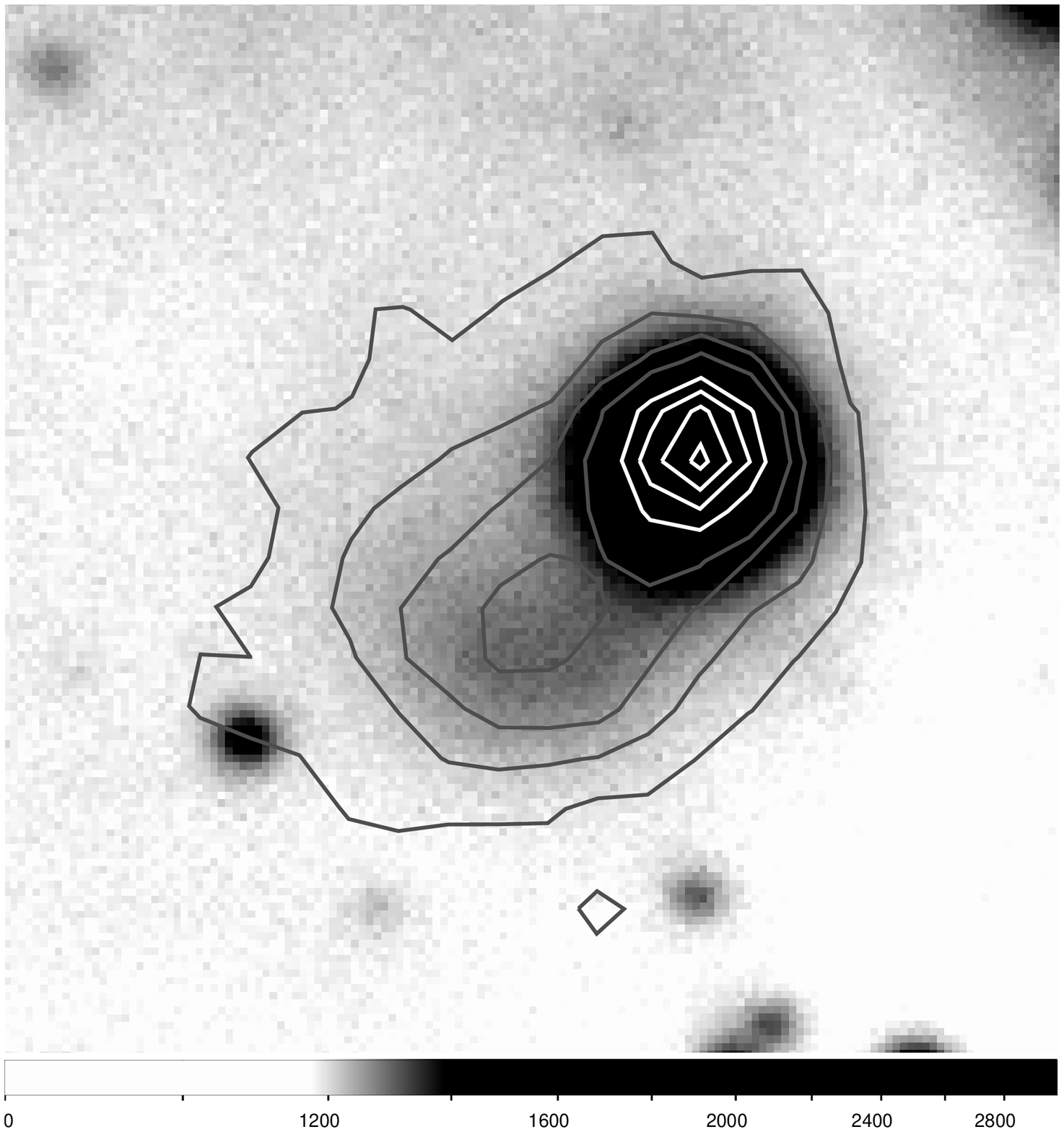}
\caption{ (left) Inner section ($8'' \times 8''$)  of the \vlt/\fors\ $R$-band image
of the Vela  Jr. CCO. North to  the top, East to the  left. The circle
($0\farcs7$  radius, $1 \sigma$) corresponds  to the recomputed \chan\  position uncertainty
after  accounting for the  accuracy of  our astrometric  solution (see
\S2.2).  Star Z of Pavlov et al.  (2001) is labelled. The cross indicates the position of the candidate CCO counterpart (see \S3.3).
(right) Isophotal contours from
the UKST  $H_{\alpha}$ image of Pellizzoni et al. (2002) overplotted.
Dark and light isophotes have a linear and geometric spacing in intensity, respectively. The coincidence of the diffuse feature 
both in position and in morphology is evident. }
\label{velajr_fors}       
\end{figure*}

\noindent
The  coordinates of  the Vela  Jr. CCO  were measured  with  a good
nominal  accuracy (0\farcs6)  by  Pavlov et  al.   (2001) who  report:
$\alpha   (J2000)$=08$^h$    52$^m$   01.38$^s$,   $\delta   (J2000)$=
-46$^\circ$  17'  53\farcs34.  We  have  re-analyzed  the same  \chan\
observation  and  we  have  obtained: $\alpha  (J2000)$=08$^h$  52$^m$
01.37$^s$, $\delta  (J2000)$= -46$^\circ$ 17' 53\farcs50,  i.e. with a
slight offset southwest but statistically consistent with the previous
ones.   

The astrometry on  the \fors\ image was computed  using as a reference
26 stars selected from the \tmass\ catalogue. The pixel coordinates of
these stars  (all non saturated  and evenly distributed in  the field)
were measured  by gaussian fitting  their intensity profiles  by using
the  specific function  of  the GAIA  (Graphical  Astronomy and  Image
Analysis)     tool\footnote{star-www.dur.ac.uk/~pdraper/gaia/gaia.html}
while the fit  to the RA, DEC reference frame  was performed using the
Starlink                                                        package
ASTROM\footnote{http://star-www.rl.ac.uk/Software/software.htm}.    The
rms of the  astrometric solution turned out to  be $\approx$ 0\farcs12
per coordinate.  After accounting for the 0\farcs2 average astrometric
accuracy of the \tmass, the  overall uncertainty to be attached to the
position of our target is finally 0\farcs65.

A $8{\arcsec}  \times 8{\arcsec}$ zoom of the \fors\ $R$-band image of the field is
shown in Fig. \ref{velajr_fors}-left.   No point-like source appears at the
\chan\  position,  thus  setting  an  upper limit  of  $R  \sim  25.6$
($3\sigma$) on  the CCO counterpart  i.e., about 3  magnitudes deeper
than  the one  obtained  by  Mereghetti et  al.   (2002).  Instead,  a
compact  optical nebula  is  cleary detected.   We  exclude that  this
nebula is an artifact  due to a PSF anomaly of star  Z, to a defect in
the image flat fielding, to ghost images, or to any other instrumental
effect.     Both    its    position    and   extent    though    (Fig.
\ref{velajr_fors}-right), are  consistent with  the one of  the $H_{\alpha}$
nebula seen by Pellizzoni et al.  (2002), which clearly indicates that
they are  the same object.  The  nebula brightness in  the $R$-band is
$\sim   22.8$   magnitude   arcsec$^{-2}$   in   the   central   part.
Unfortunately,
no color information  is available on the nebula and,  thus, it is not
possible  to  constrain  its  spectrum.   However, we  note  that  the
$R$-band  flux of  the  nebula is  $(3.7\pm0.3)\times10^{-14}$  erg
cm$^{-2}$   s$^{-1}$,  which   is   consistent  with   the  value   of
$(3.0\pm0.6)\times10^{-14}$   erg  cm$^{-2}$   s$^{-1}$   measured  in
$H_{\alpha}$  by Pellizzoni  et al.   (2002). This  suggests  that the
nebula  spectrum  has,  at  least  in the  $R$  band,  no  significant
continuum  components  and  is  dominated by  pure  $H_{\alpha}$  line
emission.

\begin{table}[t]
\begin{center}
  \caption{Summary of  the \vlt\/\naco\ $J,H,K$-band observations of  the Vela Jr.
field with the number of exposures per filter (N$\times$NDIT), the Detector Integration Time (DIT) , the average seeing and airmass  values. }
\begin{tabular}{lclllll} \hline
yyyy.mm.dd     & Fil.   &N$\times$NDIT  & DIT (s) & Seeing & Airm.	\\ \hline
2006.05.23     & $K_s$  &5$\times$19 & 24  &0\farcs78 & 1.18  \\
               & $H$    &2$\times$19 & 60  &0\farcs81  & 1.32  \\
2006.05.24     & $K_s$  &2$\times$20 & 60  &0\farcs74  & 1.19  \\
               & $J$    &2$\times$21 & 60  &0\farcs63  & 1.25  \\
\hline
\end{tabular}
\label{nacodatasummary}
\end{center}
\end{table}

\section{IR Observations}

\subsection{Observations Description}

Deep IR  observations of  the Vela Jr.  CCO were performed  in visitor
mode  on May  23rd and  24th 2006  with \nacon\  (\naco),  an adaptive
optics  (AO)  imager  and  spectrometer  mounted at  the  fourth  Unit
Telescope (UT4) of the \vlt.  In order to provide the best combination
between angular  resolution and sensitivity, \naco\  was operated with
the S27 camera with a  corresponding field of view of $28''\times28''$
and a pixel scale of 0\farcs027.  As a reference for the AO correction
we  have  used  the  \gsc\  star  S1331311130291  ($V=15.3$),  located
11\farcs3 away  from our target.  The Visual  ($VIS$) dichroic element
and wavefront  sensor ($4500-10000  \: \AA$) were  used.  Observations
were performed in the $J  (\lambda=12650 \: \AA ; \Delta \lambda= 2500
\: \AA)$, $H (\lambda=16600 \: \AA ; \Delta \lambda= 3300 \: \AA)$ and
$K_s (\lambda=21800  \: \AA ;  \Delta \lambda= 3500 \:  \AA)$ filters,
with a total net integration time  of about 2300 s per band.  To allow
for subtraction  of the variable  IR sky background,  each integration
was  split in  sequences  of short  randomly  dithered exposures  with
Detector Integration Times  (DIT) of 24 and 60  s and NDIT repetitions
along   each    point   of   the   dithering    pattern   (see   Table
\ref{nacodatasummary}).   The  instrument  readout mode  was  selected
according to the DIT in order to minimize the read out noise.  For all
 observations,  the seeing conditions  were on average  below $\sim
0\farcs8$ and the airmass was better than 1.3, allowing for an optimal
use of the  \naco\ adaptive optics. Unfortunately, in  the first night
the  AO  correction degraded  significantly  towards  the  end of  the
$H$-band exposure  sequence.  
Night (twilight flat  fields) and day time calibration frames
(darks,  lamp flat  fields) were  taken daily  as part  of  the \naco\
calibration  plan.  Standard  stars from  the Persson  et  al.  (1998)
fields were observed in both nights for photometric calibration.

\subsection{Data Reduction and Analysis}

The    data   have    been    processed   using    the   ESO    \naco\
pipeline\footnote{www.eso.org/observing/dfo/quality/NACO/pipeline} and
the  science images  reduced with  the produced  master dark  and flat
field frames.   For each band,  single reduced science  exposures have
been then combined  by the \naco\ pipeline to  produce cosmic-ray free
and  sky-subtracted  images.   The  photometric  calibration  pipeline
yielded  average zero  points of  $23.87 \pm  0.07$ ($J$),  $23.81 \pm
0.05$ ($H$), $22.93  \pm 0.03$ ($K_s$) for the  first night and $23.98
\pm 0.04$ ($J$), $23.85 \pm  0.04$ ($H$), $22.85 \pm 0.03$ ($K_s$) for
the second night.  \\

The astrometric  calibration of the \naco\ images  was performed using
the  same procedure  described  in \S2.2.   However,  since only  four
\tmass\ stars are  identified in the narrow \naco\  S27 field of view,
we have computed  our astrometry by using as a reference  a set of ten
secondary  stars  found in  common  with  our  \fors\ $R$-band  image,
calibrated  using \tmass\  (see \S2.2).   The rms  of  the astrometric
solution  turned then to  be $\approx$  0\farcs06 per  coordinate.  By
adding in quadrature the rms of the astrometric solution of the \fors\
image ($\approx$  0\farcs12 per  coordinate) and after  accounting for
the astrometric  accuracy of \tmass\, we  thus end up  with an overall
accuracy of 0\farcs37 per  coordinate on the \naco\ images astrometry.
Assuming the  same \chan\ positional  error as before  (0\farcs6), the
overall uncertainty  on the  target position on  the \naco\  images is
thus 0\farcs7.

\begin{figure}
\centering 
\includegraphics[bb=36 147 576 683,width=8.0cm,angle=0,clip]{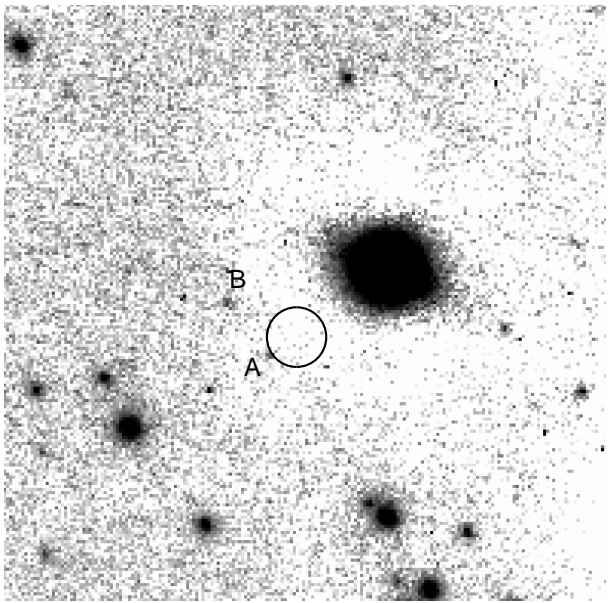}
\caption{$14''  \times  14''$  section  of   the  \vlt/\naco\  co-added
$JHK_s$-band image of the Vela Jr.  CCO. North to the top, East to the
left.   The  circle  ($0\farcs7$  radius) corresponds  to  the  \chan\
position  uncertainty  after  accounting   for  the  accuracy  of  our
astrometric solution (see \S3.2).  Star  Z of Pavlov et al.  (2001) is
visible northwest of the error circle. The faint object ($H = 21.6 \pm
0.1$; $K_s = 21.4 \pm 0.2  $), labelled A, close to the southeast edge
of  the  error circle  is  the possible  IR  counterpart  of the  CCO.
Another faint  object ($J=21.2\pm 0.2$), labelled B,  is visible $\sim
1\farcs9$ northeast from the computed \chan\ position.  }
\label{velajr_naco}       
\end{figure}

\subsection{Results}

We have  identified a very faint, apparently  point-like, object close
to the southeast edge of the  \chan\ CCO error circle (see Fig. \ref{velajr_naco}).  Although it is
detected with  a marginal significance  ($\sim 4 \sigma$),  we exclude
that this object is a  fake detection due to a background fluctuation,
to a detector feature, or  to a data reduction artifact.  Besides, its
reality  is  confirmed  by  its  independent  detection  in  both  the
$K_s$-band images  taken in the two  consecutive nights as  well as in
the  $H$-band  image. However,  the  object  is  not detected  in  the
$J$-band image and in the \fors\ $R$-band one ($R \ga 25.6$), where it
is buried by the nebula background  (see \S 2.3). Since the \naco\ PSF
is largely oversampled, we have resampled all the reduced $J$, $H$ and
$K_s$-band images to obtain a detection with a better signal-to-noise.
To  this aim,  we used  the {\em  swarp} program  (Bertin  E., Terapix
Project) to resample the \naco\ images with a $3\times3$ pixels window
while conserving the  flux.  This helped in reducing  the noise and in
enhancing  the faintest  objects. Resampled  images have  been finally
registered and  co-added.  Fig.  \ref{velajr_naco}  shows the co-added
$JHK_s$-band  image  with the  computed  \chan\  position  of the  CCO
overlaid.  The object close to  the \chan\ error circle is labelled A.
A second point-like object (labelled  B) is identified in the co-added
image,  although it  is detected  in  the $J$-band  image only,  $\sim
1\farcs9$ northeast  from the computed \chan\  position.  However, the
almost $2  \sigma$ offset from the  edge of the  error circle suggests
that it is likely a background object. We thus assume that object A is
the only possible counterpart to the Vela Jr. CCO.

We   have  computed   the   magnitudes  of   object   A  through   PSF
photometry.  Although the accuracy of the PSF photometry certainly
depends  on  the  AO   correction,  the  target  faintness  makes  PSF
photometry more  suitable than standard aperture  photometry. To this
aim, we  have used  the Stetson (1992,  1994) suite of  programs \dao,
following  the procedures  outlined in  Zaggia et  al.  (1997).   As a
reference for our  photometry we have used the  resampled $J$, $H$ and
$K_s$-band images, which have a much better signal-to-noise.  For each
image we then calculated the model PSF with \dao\ using a 6 parameters
Penny  function   to  fit  the  profile  of   the  $\sim15$  brightest
non-saturated objects  in the field.  The  model PSF was  then used to
measure the fluxes of object A as  well as of all other objects in the
field.  To improve the precision of  our PSF photometry, as well as to
maximize the number  of objects identified in the  field, we have used
the \all\  routine of \dao.   From the co-added $JHK_s$-band  image we
have  run \dao\  to create  a  master list  of objects  which we  have
properly registered on each single band image and then passed to \all\
to  be used  as a  reference  for PSF  photometry in  the single  band
images.  The single band  photometry catalogues have been then matched
by  \all\  to  create   the  final  multi-band  photometry  catalogue.
Photometry  measurements obtained  in the  two $K_s$-band  images have
been averaged.   We performed  the absolute photometry  calibration of
our PSF photometry  using the zero points provided  by the \naco\ data
reduction pipeline. Since the  \naco\ zero points are computed through
aperture  photometry,  the  magnitudes  computed by  \dao\  have  been
corrected by applying the aperture correction.  We finally applied the
airmass correction  using the atmospheric  extinction coefficients for
the Paranal Observatory.  The magnitudes of object A then turns out to
be $H = 21.6  \pm 0.1$ and $K_s = 21.4 \pm 0.2  $. Apart from object B
($J=21.2\pm  0.2$), no other  object is  detected within/close  to the
\chan\ error circle down to $J \sim 22.6$, $H \sim 22.5$ and $K_s \sim
21.8 $.

\section{Discussion}

\subsection{The nebula}

 Clumpy emission structures  and filaments produced by shocks from
the supernova  explosion are detected  in $H_{\alpha}$ in  young SNRs,
normally (but not only) along the external rims. Thus, it is possible,
although very unlikely, that the  CCO projected position is by chanche
coincident with one  of these structures.  More likely,  the nebula is
produced by the neutron star ionization of the neutral Hydrogen in the
SNR. We  note that  the existence of  neutral Hydrogen in  young SNRs,
following atoms recombination after  the supernova explosion, has been
confirmed in a number of cases (see, e.g. Michael 2000).  

One  of the  possible  interpretations of  the  nebula (Pellizzoni  et
al. 2002)  is in  terms of  a bow-shock produced  by the  neutron star
motion through  the ISM.  Interestingly, the $R$-band  image shows for
the  first time  that the  nebula is  not spherically  symmetrical but
looks  more like a  kidney bean,  a shape  somehow reminescent  of the
arc-like  structures observed  around bow-shocks.   Unfortunately, the
spatial resolution  of the  \vlt\ image is  not sufficient  to resolve
more clearly the  nebula morphology and to map  its surface brightness
distribution which, in the  case of a velocity-driven bow-shock, would
present a sharp enhancement towards  the apex, depending on the actual
value and  direction of the neutron  star velocity.  We  note that the
position of  the CCO lies behind  the apparent apex of  the nebula and
roughly on its projected axis  of simmetry, right where expected for a
velocity-driven bow-shock.   If object A were indeed
the CCO counterpart, its position would  be quite close to
the nebula and slightly displaced  from its apparent projected axis of
symmetry.   However, given  the  poor characterization  of the  nebula
morphology and  the lack  of information on  the neutron  star spatial
velocity,  this  would  not  affect  the validity  of  the  bow  shock
interpretation.

Since  no proper  motion measurement  is  yet available  for the  Vela
Jr. CCO,  the bow-shock interpretation can not  be directly confirmed.
No counterpart of  the putative bow-shock has been  detected so far in
radio and  in the X-rays.   However, we note  that the detection  of a
bow-shock in $H_{\alpha}$ only would  not be in contradiction with its
non-detection  at  other  wavelengths.   Indeed, the  detection  of  a
bow-shock  in  $H_{\alpha}$  is  typically  anti-correlated  with  its
detection  in  radio and  in  the  X-rays,  with the  only  noticeable
exception of the bow-shock around the ``Black Widow'' pulsar (Stappers
et   al.   2003).    A   possible  argument   against  the   bow-shock
interpretation comes from the position  of the CCO with respect to the
estimated  geometrical center  of  the Vela  Jr.   SNR.  According  to
Aschenbach (1998),  this should be located  at $\alpha (J2000)$=08$^h$
52$^m$ 03$^s$,  $\delta (J2000)$= -46$^\circ$  22', which is  $\sim$ 4
arcmin south  of the  CCO position. This  would imply a  proper motion
direction different from the one expected from the apparent morphology
of the  nebula, which  would be more  compatible with a  proper motion
vector  pointing  southwest.  However,  both  the  uncertainty in  the
definition  of  the geometrical  center  of  the  SNR, see  e.g.   the
representative case of the Vela SNR (Aschenbach et al.  1995), and the
still poorly characterized morphology of the nebula do not make this a
strong  argument  to  rule   out  the  bow-shock  interpretation.   

An alternative possibility is that  the nebula around the Vela Jr. CCO
is a photo-ionization nebula, powered by the UV radiation from the hot
neutron  star. The  possible  structure of  a photo-ionization  nebula
associated  to  a  neutron  star  moving  supersonically  through  the
interstellar  medium has  been  studied by  Van  Kerkwijk \&  Kulkarni
(2001).   According to  their model,  a velocity-driven  morphology is
expected, with  a definite symmetry  with respect to the  direction of
the projected space velocity of the NS. Such a nebula should also have
a  smooth brightness  profile,  rather  different from  the  one of  a
bow-shock. Our VLT  $R$-band image points to a  morphology which could
be consistent  with such a  picture. However, as  for the case  of the
bow-shock scenario, any firm conclusion  is prevented by the non sharp
enough angular resolution of the ground-based images.

 Altough most of  the nebula emission is in  $H_{\alpha}$ (see \S 2.2),
we can not a priory rule  out the possibility of a synchrotron nebula,
powered by the relativistic particle wind emitted by the neutron star.
These   pulsar-wind   nebulae   are   often  observed   around   young
rotation-powered pulsars, usually in the  radio and in the X-ray bands
but also in the optical band.  Bright optical pulsar-wind nebulae have
been indeed observed around the Crab pulsar (e.g.  Hester et al. 1995)
and  PSR  B0540--69 (Caraveo  et  al.   2001).   However, while  these
objects also  feature X-ray  pulsar-wind nebulae, with  structures and
morphologies  very similar  to the  optical ones,  no evidence  for an
X-ray nebula  has been found  for the Vela Jr.   CCO.  High-resolution
\chan\ X-ray observations (Pavlov et  al.  2001) shows that the source
profile is consistent  with the instrument's PSF, which  rules out the
presence  of an extended  emission component,  unless it  is extremely
faint.   Recent  radio  observations  performed  with  the  Australian
Telescope Compact Array (Reynoso et al.  2006) have also ruled out the
presence  of a  pulsar-wind nebula  in radio.   Thus, the  lack  of an
X-ray/radio counterpart, together with  the apparent lack of continuum
optical emission, makes the  pulsar-wind nebula interpretation for the
Vela Jr.  nebulosity very unlikely.

\begin{figure}
\centering 
\includegraphics[width=8.5cm,angle=0]{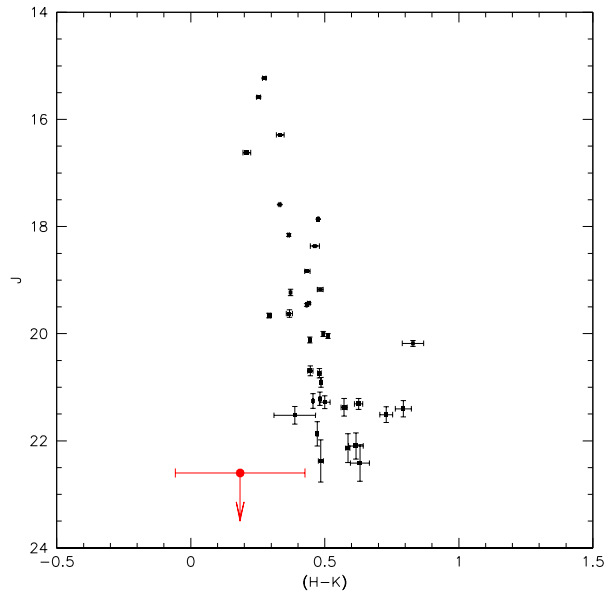} 
\caption{(left) $J$ vs $H-K_s$ color  magnitude diagram built for all objects
 detected  in the  $\approx 30''  \times 30''$  field around  the Vela
 Junior CCO.  Field objects are  marked by filled black  squares.  The
 possible counterpart  to  the  CCO  is marked  by  the  filled  red
 circle. The horizontal bar  corresponds to the possible $H-K_s$ range
 derived from the attached photometry errors. }
\label{velajr_cmd}       
\end{figure}

\subsection{The point source}

At the  moment we  can not confidently  rule out the  possibility that
object  A is  just  a field  object  unrelated to  the  CCO, whose  IR
counterpart would thus remain unidentified.  Indeed, given the density
of stars in the \naco\ field  of view we estimate a chance coincidence
probability of $\sim 15\%$,  i.e. certainly not negligible.  Given our
flux upper  limits ($R \sim  25.6$, $J \sim  22.6$, $H \sim  22.5$ and
$K_s  \sim  21.8   $),  this  would  almost  certainly   rule  out  an
hypothetical  stellar  companion  (see  discussion  below)  and  would
instead support  the conclusion  that the CCO  is an  isolated neutron
star, possibly surrounded by a circumstellar disk.  However, if object
A is indeed the CCO  counterpart, its nature is controversial.  As the
matter of  fact, it is  not a priori  obvious whether the  detected IR
emission can  be ascribed  to the neutron  star itself, to  a fallback
disk  around  the neutron  star,  or  to  a putative  companion  star.
Suggestive  as it  could be,  the object  does not  show  evidence for
variability,  with the  $K_s$ magnitudes  measured in  the  two nights
consistent within the associated errors.

The first possibility is that object A is itself the IR counterpart of
the neutron star.  So far, IR emission has been detected only for five
rotation-powered pulsars (see Mignani et  al. 2007b for a summary) and
it  is  ascribed to  synchrotron  radiation  produced by  relativistic
particles  in  the  neutron  star's magnetosphere.   The  observed  IR
luminosity spans about six decades and it is strongly dependent on the
neutron star's age. If we assume for  the Vela Jr. CCO an age of a few
thousands years, as estimated for the  host SNR, it might then have an
IR luminosity  comparable to the one  of the Vela  pulsar, i.e.  $\sim
10^{28}$ ergs s$^{-1}$.   At a distance of 1 kpc it  would then have a
magnitude $K_s  \sim24.1$, i.e.  much  fainter than the one  of object
A. IR  emission from  an isolated neutron  would thus  be incompatible
with our  detections unless the CCO  distance is as low  as $\sim 300$
pc,  or  its  IR  luminosity   is  a  factor  $\sim  10$  higher  than
expected. However,  given both the  uncertainties on the  CCO distance
and the  lack of information  on its energetics,  we can not  a priori
exclude that object A is itself the counterpart of the neutron star.

An alternative  possibility is that the  IR emission from  object A is
produced from a fallback disk around the neutron star.  Fallback disks
have also  been invoked to explain  the IR spectral  flattening of the
AXPs (e.g.,  Israel et al.  2003;  Israel et al.  2004)  and have been
searched around other objects  possibly related to the magnetars, like
the high-magnetic field radio pulsars  (Mignani et al.  2007b) and the
XDINSs  (Lo   Curto  et  al.    2007)  but,  so  far,   with unconclusive
results. Evidence  for a  fallback disk has  been possibly  found from
\spitzer\ observations of the AXP  4U\,0142+61 (Wang et al.  2006). We
note that the  $F_{IR}/F_X$ ratio of the Vela  Jr. CCO is $\sim5\times
10^{-4}$, which is very similar to  the value observed for the case of
AXP 4U\,0142+61. Accurate multiband IR photometry is required in order
to test such an hypothesis.

The last possibility is that object A is the stellar mass companion of
the CCO.  We have tried to determine a possible stellar classification
of object A by  using the available color information.  Unfortunately,
its non-detection  in the $J$  band leaves us  with only a  measure of
$H-K_s$ and with upper limits on $J-H$ and $J-K_s$.  However, the fact
that object A  has not been detected  in the $J$ band but  only in the
$H$  and $K_s$  bands, despite  of similarly  deep exposures,  seem to
imply that its spectrum is  quite red.  To better evaluate its colors,
we have  compared its  location in a  $J$ vs. $H-K_s$  color magnitude
diagram (CMD) with  respect to those of a number  of stars detected in
the field  (Fig.  \ref{velajr_cmd}a).   Photometry of the  field stars
has been computed  as described in \S 3.3.  Unfortunately, only $\sim$
50 stars  have been  identified in the  narrow and  sparsely populated
\naco\ field of view,  which certainly hampers our characterization of
the field star sequence. At a  first glance, with an $H-K_s \sim 0.2$,
object  A indeed  seems to  deviate from  the sequence.   However, its
large $H-K_s$ error makes this deviation not significant. Thus, object
A is  not redder, possibly only  slightly bluer, than  the majority of
the field stars.  This can  be visually appreciated from the composite
$JHK_s$-band  image of  the field  (Fig.  \ref{velajr_cmd}b).   We have
investigated the  effects of the  interstellar extinction on  the CMD.
The interstellar reddening towards the  Vela Jr.  CCO can be estimated
from the Hydrogen column density $\nh \approx 3.5 \: 10^{21}$ cm${-2}$
derived  from  the  X-ray  observations  (Kargaltsev  et  al.   2002).
According to the  relation of Predhel \& Schmitt  (1995) this gives an
$E(B-V) \sim 0.67$ which corresponds to $A_J \sim 0.5$, $A_H \sim 0.3$
and $A_K \sim 0.2$ assuming the extinction coefficients of Fitzpatrick
(1999).   However,  accounting  for  the  reddening  does  not  affect
significantly the location of object A  in the CMD with respect to the
sequence of  the field stars. Thus,  we conclude that object  A has no
peculiar colors  and it  does not stand  out from the  average stellar
population in the field.

The observed colors of object A,  $H-K_s =0.2 \pm 0.2$, $J-H \ge 1.0$,
and $R-K_s \ge 5.2$ can be compatible with those of a mid M type star,
after accounting  for the  interstellar reddeding (e.g.  Legget 1992).
For instance, an  M5 star at a  distance of 2 kpc would  be just below
our  detection  limit  in  $J$  but  would  have  $H$  and  $K_s$-band
magnitudes consistent  with the ones  measured for object A.   We note
that a  distance of  2 kpc  is larger than  the one  of 1  kpc usually
assumed for the SNR (Slane et al.  2001), but is not incompatible with
the observations.   Thus, our  data are consistent  with the  Vela Jr.
CCO being a  neutron star in a binary system with  an M type companion
star.  However, since we have only  a lower limit on $J-H$, we can not
exclude that  object A is  actually a star  of an even  later spectral
type, an example  of the recently discovered L  dwarfs (Kirkpatrick et
al. 1999).  Late type star counterparts have been also proposed in the
past for the  PKS 1209-52 (Pavlov et al.  2004)  and RCW103 (Sanwal et
al.   2002) CCOs but  both identifications  have been  later discarded
from a  re-analysis of the  \chan\ astrometry (Mignani et  al.  2007a;
Wang et al.  2007; De Luca et al. private communication).  If our identification is
confirmed, the association of a  young neutron star in a binary system
with such  a low  mass companion will  pose more than  one theoretical
challenge for  current neutron star formation  and supernova explosion
models.

\section{Conclusions}

We  have  reported  on  optical/IR  \vlt\  observations  of  the  Vela
Jr. CCO. The \vlt\ $R$-band data have revealed a compact nebula at the
\chan\  position of  the CCO,  consistent  both in  morphology and  in
brightness with  the $H_{\alpha}$  nebula identified by  Pellizzoni et
al.   (2002). We  can not  presently determine  whether the  nebula is
associated to  a bow-shock  produced by the  supersonic motion  of the
neutron star in  the ISM, or it is  a photo-ionization nebula produced
by the UV  thermal radiation from the neutron  star. Higher resolution
imaging is required  to resolve the nebula structure  and to determine
whether   it  is   more   compatible  with   the   bow-shock  or   the
photo-ionization  scenario,   for  which  we   expect  quite  distinct
morphologies,    i.e.    arc-like    or    spherical,    respectively.
Alternatively, high-resolution spectroscopy  would make it possible to
measure the width of the  $H_{\alpha}$ line and to derive its velocity
broadening which  is typically higher for a  velocity driven bow-shock
nebula.

Proper  motion  measurement  of  the  CCO also  represent  a  powerful
diagnostic to investigate the  bow-shock scenario, where we expect the
velocity vector  to lie along the  simmetry axis of the  nebula and to
point towards its apex.  Since the CCO is radio quiet, one possibility
to obtain a proper motion measurement would be through high resolution
X-ray  observations with  \chan.  However,  proper  motion measurement
performed with  \chan\ on some nearby  INSs (Motch et  al.  2007) only
allowed  to  obtain  marginally  significant results.   For  the  Vela
Jr. CCO which  is up to a factor five  more distant, such measurements
are  even  more  difficult.    More  realistically,  a  proper  motion
measurement could  be obtained from high-resolution IR  imaging of the
candidate CCO  counterpart identified in our \vlt\  $H$ and $K_s$-band
observations.  Its  proper motion measurement  will allow to  test the
bow-shock interpretation  for the nebula  on one hand, and  to confirm
its proposed identification, which only relies on the coincidence with
the \chan\ position, on the other.

The  nature   of  the  candidate  CCO  counterpart   is,  at  present,
unclear. It could be the neutron  star itself, or a fallback disk left
over  from the  supernova explosion,  or a  very late  type star  in a
binary system  with the  neutron star.  In  all cases,  confirming the
identification  will  have  several  important  implications  for  our
understanding  of the  CCOs  as class,  as  well as  for neutron  star
formation  and supernova  explosion models.   Instead,  discarding the
proposed  identification will strengthen  the more  standard framework
isolated neutron star scenario for the CCOs.

\begin{acknowledgements}

RPM warmly thanks N. Ageorges (ESO) for her invaluable help and friendly support at the telescope,  and D.   Dobrzycka (ESO) for  reducing the IR data with  the \naco\ pipeline.

\end{acknowledgements}

\end{document}